\documentclass[fleqn,usenatbib]{mnras}

\usepackage{newtxtext,newtxmath}
\usepackage[T1]{fontenc}
\usepackage{graphicx}
\usepackage{amsmath,scalerel}
\usepackage{relsize}
\usepackage{subcaption}
\usepackage{soul}
\usepackage{booktabs}
\usepackage{comment}
\usepackage{ulem}

\DeclareRobustCommand{\VAN}[3]{#2}
\let\VANthebibliography\thebibliography
\def\thebibliography{\DeclareRobustCommand{\VAN}[3]{##3}\VANthebibliography}

\definecolor{sk}{rgb}{0.99, 0.00, 0.5}

\title[Morphology from massive to dwarf galaxies]{Global trends in morphology from massive to dwarf galaxies}

\author[I. Lazar et al.]{I. Lazar\thanks{E-mail: i.lazar@herts.ac.uk},$^{1}$ S. Kaviraj,$^{1}$ C. J. Conselice,$^{2}$ L. Westcott,$^{2}$ A. E. Watkins,$^{1}$ S. Koudmani,$^{1,3}$ G. Martin,$^{4}$ \newauthor T. M. Sedgwick,$^{1}$ D. Kakkad$^{1}$ and B. Bichang'a$^{1}$\\ 
$^{1}$Centre for Astrophysics Research, Department of Physics, Astronomy and Mathematics, University of Hertfordshire, College Lane, Hatfield AL10 9AB, UK\\
$^{2}$Jodrell Bank Centre for Astrophysics, University of Manchester, Oxford Road, Manchester M13 9PL, UK\\
$^{3}$St Catharine's College, University of Cambridge, Trumpington Street, Cambridge CB2 1RL, UK\\
$^{4}$School of Physics and Astronomy, University of Nottingham, University Park, Nottingham NG7 2RD, UK
}

\pubyear{\the\year{}}

\begin{document}
\label{firstpage}
\pagerange{\pageref{firstpage}--\pageref{lastpage}}
\maketitle

\begin{abstract}
The morphological properties of dwarf galaxies ($M_{\star}$ $<$ 10$^{9.5}$ M$_{\odot}$) remain largely unexplored, particularly outside the local neighbourhood. We explore how morphology changes across the massive to dwarf-galaxy regimes, using a mass-complete sample of $\sim$1000 galaxies, with stellar masses and redshifts in the ranges 10$^{7}$ M$_{\odot}$ $<$ $M_{\star}$ $<$ 10$^{12}$ M$_{\odot}$ and $z$ $<$ 0.15 respectively. By combining JWST-derived morphological parameters (concentration, asymmetry and clumpiness; `CAS') and visual morphological classifications, we explore: (1) how morphology changes with stellar mass and effective surface brightness, (2) the connection between morphology and recent star formation history, as a function of stellar mass, (3) how bar frequency changes between the massive and dwarf regimes and (4) how well the CAS parameters perform in separating early- and late-type galaxies, as a function of stellar mass. We demonstrate that galaxies become less concentrated, more asymmetric and less clumpy with decreasing stellar mass. In both mass regimes, galaxies that are more concentrated and less asymmetric are more likely to be red (i.e. quenched). The decrease in concentration towards lower stellar masses results in a loss of the leverage that this parameter can provide in separating early- and late-type galaxies. Thus, while the CAS system successfully separates early- and late-type systems in the massive-galaxy regime, these morphological classes become significantly more difficult to separate, using these parameters, in the dwarf regime. Finally, the bar fraction declines steadily with decreasing stellar mass and becomes consistent with zero at $M_{\star}$ $\sim$ 10$^{8}$ M$_{\odot}$, suggesting a lower limit for the galaxy mass needed to induce bar formation.
\end{abstract}

\begin{keywords}
galaxies: formation -- galaxies: evolution -- galaxies: dwarf -- galaxies: star formation -- galaxies: structure -- galaxies: fundamental parameters
\end{keywords}


\section{Introduction}

Morphology is a fundamental quantity in observational astrophysics. In at least the massive-galaxy ($M_{\star}$ > 10$^{10}$ M$_{\odot}$) regime, morphology has been shown to correlate with key characteristics of a galaxy, like its stellar mass \citep[e.g.][]{Bundy2005,Buitrago2014}, star formation rate \citep[SFR, e.g.][]{Whitaker2015}, rest-frame colour \citep[e.g.][]{Conselice2006}, merger history \citep[e.g.][]{Martin2018_sph} and local environment \citep[e.g.][]{Dressler1980,Dressler1997,Skibba2009}. The morphological properties of galaxies therefore encode details of their formation histories, making them a valuable tool for understanding how the observable Universe evolves over its lifetime. However, the current morphological literature is overwhelmingly dominated by massive galaxies, largely because these are the systems that are bright enough to be detected in past survey across large fractions of cosmic time. In contrast, as we discuss further below, the detection limits of past surveys make it difficult to construct complete, unbiased dwarf populations outside the very local Universe. As a result our understanding of how morphological properties change from the massive to the dwarf regime, particularly in low-density environments which host the majority of galaxies, remains a key open question. It is worth noting here that studying morphology in the dwarf regime may provide deeper insights into the drivers of galaxy evolution than their counterpart studies in massive galaxies \citep[e.g.][]{Martin2025,Watkins2026}, because the shallow potential wells of dwarfs make their internal structure more susceptible to being perturbed by a wide variety of physical processes \citep[e.g.][]{Aarseth1972,Goodwin2004,Bastian2011,Lelli2014,Kimbrell2021,Watkins2025}. 

Given their intrinsic faintness, past dwarf studies that have employed mass-complete populations have typically focussed either on the Local Volume or, due to the lack of precise distance estimates, have relied on associating dwarfs in the very local Universe with a more massive structure such as a host galaxy, group, or cluster \citep[e.g.][]{Tolstoy2009,Duc2015,Geha2017,Venhola2018,Trujillo2021,Poulain2021,Mao2021,Watkins2023}. While statistical studies of dwarfs in low-density environments outside the local neighbourhood do exist, many of these papers utilise surveys like the SDSS \citep{Alam2015} which, notwithstanding their large footprints, are relatively shallow. As discussed in detail in \citet{Kaviraj2025}, typical dwarfs are too faint to be detectable in shallow surveys outside the local neighbourhood. The dwarfs that do appear in such datasets have anomalously high SFRs, because the (brighter) young stellar populations are needed to boost the luminosity of the dwarfs above the detection thresholds of shallow surveys like the SDSS. Consequently, the dwarf populations in these surveys are both incomplete and strongly biased towards blue systems, which makes it challenging to draw accurate conclusions about the dwarf population as a whole\footnote{Some studies using the SDSS do contain galaxy populations (including dwarfs) which are almost complete but these studies trace very low ($z<0.01$) redshifts \citep[e.g.][]{Ann2015}.}. 

As a result, how galaxy morphology and its relationship with other key physical parameters changes as we transition from the massive to dwarf galaxy regime remains a key open question. For example, how does morphology vary as a function of stellar mass and effective surface brightness across these two regimes? Does the connection between morphology and star formation history (traced using rest-frame colour) change as we move from massive to dwarf galaxies? How does the incidence of internal structures like bars evolve with stellar mass? How well do quantitative morphological parameters perform when separating physical morphological classes like `early-type' and `late-type' galaxies, as we move from the massive to the dwarf regime? 
Some insights into these questions can be gained from the existing dwarf literature, albeit via small galaxy samples which are, in many cases, restricted to relatively high-density environments. For example, \citet{Poulain2021} and \citet{Carlsten2021} show that dwarfs with lower stellar masses tend to exhibit larger axial ratios, indicating that they are thicker, puffier, and more spherical (i.e. closer to oblate spheroids) than their more massive counterparts. {\color{black}This generally leads to both the Sersic index (which is a measure of the central light concentration; \citealt{Sersic1963}) and the effective radius exhibiting smaller values in dwarfs than in massive galaxies, and the Sersic index being roughly independent of stellar mass for M$_\star$ $<$ 10$^{9}$ M$_\odot$ (see e.g. Fig 7 in \citealt{Carlsten2021}).} 
 
In a similar vein, a comparison of the morphological properties of nearby ($z<0.08$) dwarfs and massive galaxies, in unbiased mass-complete samples in low-density environments \citep[e.g.][see also \citealt{Conselice2003}]{Lazar2024a} suggests that both concentration and clumpiness tend to decline with decreasing stellar mass. \citet{Lazar2024b} demonstrate, albeit with ground-based images, that the decrease in concentration at lower stellar masses leads to a reduction in the leverage that this quantity provides in separating visual morphological classes like early-type and late-type galaxies. As a result, it may become more difficult to separate visual morphological classes using traditional morphological parameters as we probe deeper into the dwarf regime. Finally, while some studies in the literature indicate that the bar fraction may decline as stellar mass decreases \citep[e.g.][]{Geron2021}, a definitive conclusion requires an exploration of bar fractions down into the dwarf regime using mass-complete galaxy samples. 

The study of dwarf galaxies is poised to be revolutionised by the advent of new and forthcoming datasets which are both deep and wide, such as the Legacy Survey of Space and Time \citep[LSST, e.g.][]{Ivezic2019} and surveys from \textit{Euclid} \citep{Laureijs2010,Mellier2025}. These datasets will offer mass-complete dwarf samples of unprecedented size, which will enable novel studies of dwarf galaxy evolution in the nearby Universe. However, deep photometric data, coupled with high resolution imaging, already exists in small regions of the sky, like the 2 deg$^2$ COSMOS field, which can provide insights into the questions posed above and offer a preview of the cutting-edge science that will soon become possible using the LSST and \textit{Euclid}. Indeed, the shape of the mass function \citep[e.g.][]{Wright2017} ensures that statistically significant samples of dwarfs can be constructed in fields with modest areas like COSMOS. Here, we perform a first, broad comparison of the morphological properties of dwarfs to their massive counterparts, using a mass complete sample of galaxies in the nearby Universe. The use of high resolution images from the James Webb Space Telescope (JWST) enables us to probe down to $M_{\star}$ $\sim$ 10$^{7}$ M$_{\odot}$ and out to $z\sim0.15$. 


The plan for this paper is as follows. In Section \ref{sec:data}, we describe the galaxy sample that underpins this study, the calculation of morphological (CAS) parameters and the visual classification of galaxy morphologies and identification of barred galaxies. In Section \ref{sec:morphology_mass}, we explore how galaxy morphology changes as a function of stellar mass and effective surface brightness. In Section \ref{sec:CAS_recent_sfh}, we explore the relationship between morphology and recent star formation history (traced via rest-frame colour) across the massive and dwarf regimes. In Section \ref{sec:CAS_to_morphology}, we study how well early-type and late-type galaxies can be separated, using classical morphological parameters, as stellar mass changes. Finally, in Section \ref{sec:bar_fraction}, we quantify the incidence of bars in the dwarf galaxy regime. We summarize our findings in Section \ref{sec:summary}. 

\begin{figure*}
\includegraphics[width=0.9\textwidth]{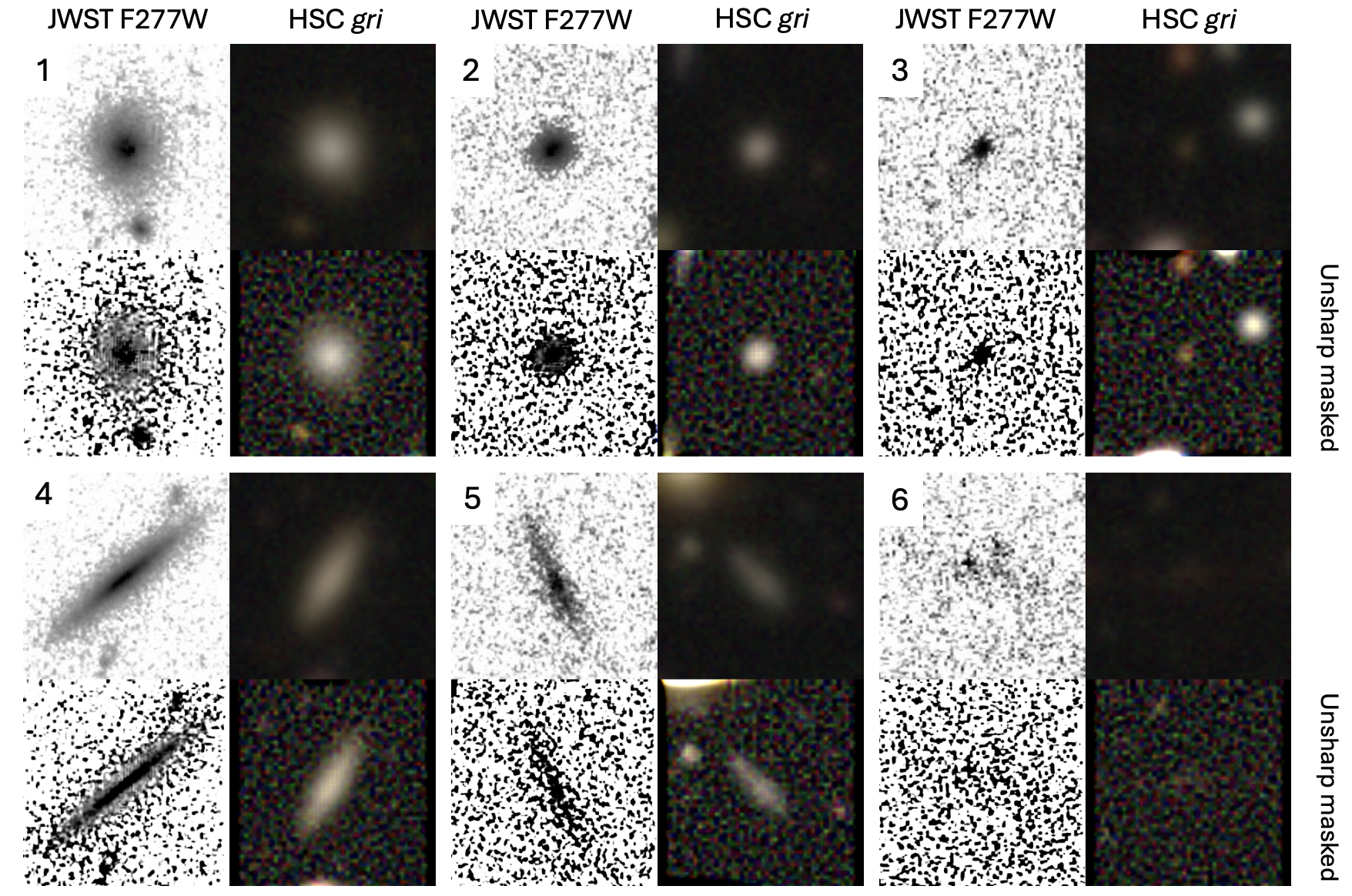}
\caption{Example images of ETGs (1, 2), galaxies classified as compact (3), LTGs (4, 5) and those flagged as unclassifiable (6). The size of each image is 5 arcseconds on a side. The images from different instruments have different orientations -- we intentionally keep it this way so that each galaxy is visually inspected at different orientations. In each image the first and second columns show the JWST and HSC images respectively. Each individual image has four panels. Within each image the panels in the first row show the original images from the different instruments, while the second row shows their unsharp-masked counterparts.} 
\label{fig:example_galaxy_images}
\end{figure*}


\section{Data}
\label{sec:data}

We combine physical parameters (stellar masses, photometric redshifts and rest-frame colours) from the COSMOS2020 catalogue \citep{Weaver2022} with structural information derived via morphological parameters and visual inspection of JWST images in the F277W filter \citep{Adams2024,Conselice2025} and those from the Hyper Suprime-Cam (HSC) in the $g$, $r$ and $i$-band filters. The HSC data is taken from the third data release \citep{Aihara2022} of the HSC Subaru Strategic Program (HSC-SSP). For our purposes we create HSC $gri$ colour-composite images using the Python function \texttt{make\_lupton\_rgb} \citep[described in][]{Lupton2004} from the Python library \texttt{astropy}. The widths of the point spread function in the JWST F277W and HSC $i$ band images are $\sim$0.1 and $\sim$0.6 arcseconds respectively. In the following sections we describe the properties of the datasets used here.  


\subsection{A mass-complete dwarf galaxy sample from the COSMOS2020 catalogue}
\label{sec:C2020_sample}

Our sample of galaxies is constructed using the Classic version of the COSMOS2020 catalogue, which provides physical parameters -- e.g. stellar masses and photometric redshifts -- for around 1.7 million sources in the $\sim$2 deg$^2$ COSMOS \citep{Scoville2007} field (centered at 10h, +02$^{\circ}$). The parameters are derived via the \textsc{LePhare} SED-fitting algorithm \citep{Arnouts2002,Ilbert2006} using photometry in around 40 broad and medium band filters spanning the UV through to 4.5 $\mu$m from the following instruments: GALEX \citep{Zamojski2007}, MegaCam/CFHT \citep{Sawicki2019}, ACS/HST \citep{Leauthaud2007}, Hyper Suprime-Cam \citep{Aihara2019}, Subaru/Suprime-Cam \citep{Taniguchi2007,Taniguchi2015}, VIRCAM/VISTA \citep{McCracken2012} and IRAC/Spitzer \citep{Ashby2013,Steinhardt2014,Ashby2015,Ashby2018}. 

The detection image incorporates optical ($i,z$) images from the Ultradeep layer of the HSC-SSP, which have point-source depths of $\sim$28 mag \citep{Aihara2019} and are $\sim$10 mag deeper than the magnitude limit of the SDSS spectroscopic main galaxy sample \citep[e.g.][]{Alam2015}. Optical and infrared aperture photometry are extracted using the \textsc{SExtractor} \citep{Bertin1996} and \textsc{IRACLEAN} \citep{Hsieh2012} codes respectively. The deep data, which straddles a large wavelength baseline, yields photometric redshift accuracies better than $\sim$1 and $\sim$4 per cent for bright ($i<22.5$ mag) and faint ($25<i<27$ mag) galaxies respectively. Here, we use measured stellar masses, photometric redshifts, rest-frame colours and SFRs directly from the COSMOS2020 catalogue. 

To reduce the impact of biases in our results it is important to work with a galaxy sample that is mass complete. In other words, we wish to use a sample that, at a given stellar mass, includes all objects, regardless of their star formation history. To construct such a mass-complete sample, we follow the methodology of \citet{Kaviraj2025}, who {\color{black}estimate the redshift} out to which a population of galaxies at a given stellar mass is likely to be complete in the COSMOS2020 catalogue. 

This redshift is defined as that at which a purely-old `simple stellar population' (SSP) of a given stellar mass, that forms in an instantaneous burst at $z=2$, is detectable, at the depth of the HSC Ultradeep imaging (which underpins object detection in COSMOS2020). This purely-old SSP is considered to be a faintest `limiting' case, since real galaxies (which are not composed solely of old stars), will be more luminous than this limiting value. Thus, if this limiting case is detectable at the depth of a given survey, then the entire galaxy population (at a given stellar mass) should also be detectable in that survey.

The top panel of Figure 4 in \citet{Kaviraj2025} uses this methodology to quantify the redshifts at which various surveys are complete. {\color{black}Since object detection in the COSMOS2020 is underpinned by the HSC UltraDeep data, the HSC UltraDeep curve corresponds to the completeness thresholds for our dataset.} This curve indicates that the galaxy population in COSMOS2020 is likely to be complete down to stellar masses of 10$^{7}$ M$_{\odot}$, out to at least $z\sim0.15$. The JWST survey area covers 0.6 deg$^2$ within the COSMOS2020 footprint. Since we require morphological data from JWST, our analysis is based on the $\sim$1000 COSMOS2020 galaxies with $M_{\star}$ > 10$^{7}$ M$_{\odot}$ with redshifts less than $z=0.15$, which lie within the JWST footprint. 

To compare the morphological properties of our dwarfs to those seen in massive galaxies, we also construct a mass-complete massive-galaxy sample which consists of $\sim$700 COSMOS2020 galaxies with $M_{\star}$ > 10$^{10}$ M$_{\odot}$ and $z<0.4$ that reside within the JWST footprint. Note that the larger redshift range in the massive-galaxy regime is needed to construct a statistically significant sample of massive galaxies, because the number density of massive galaxies is much lower than that of their dwarf counterparts \citep[e.g.][]{Wright2017}. As we discuss in Section \ref{sec:CAS_parameters} below, there is no evolution in the morphological parameters of massive galaxies in this redshift range. 

Finally, it is worth exploring the types of large-scale structures that are likely to exist within our COSMOS2020 footprint, which is not centred on a known region of high density. We consider the \textit{M}$_{200}$ values, which are proxies for the virial masses, of groups identified in the literature \citep{Finoguenov2007,George2011,Gozaliasl2014,Gozaliasl2019}, within our COSMOS footprint at our redshifts of interest. These virial masses lie in the range 10$^{12.9}$ M$_{\odot}$ < \textit{M}$_{\rm{200}}$ < 10$^{13.8}$ M$_{\odot}$, with a median value of 10$^{13.4}$ M$_{\odot}$. For comparison, a small cluster like Fornax has a virial mass of $\sim$10$^{13.9}$ M$_{\odot}$ \citep{Drinkwater2001}, while large clusters like Virgo and Coma have virial masses of $\sim$10$^{15}$ M$_{\odot}$ \citep[e.g.][]{Fouque2001,Gavazzi2009}. The galaxy population considered in this study therefore spans an array of low-density environments, from relatively large groups to the field, but does not reside in rich clusters, in line with the conclusions of the recent literature \citep[e.g.][]{Martin2025}.  

\begin{figure*}
\includegraphics[width=0.375\textwidth]{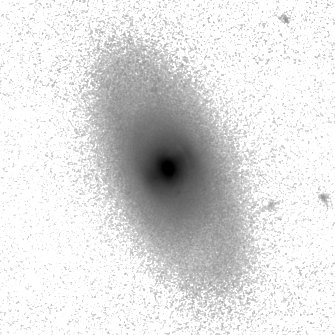}
\includegraphics[width=0.5\textwidth]{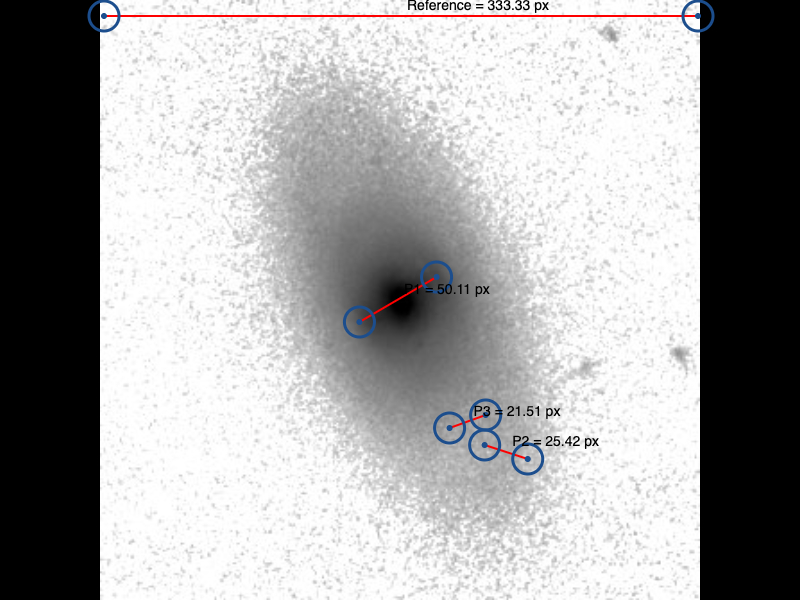}\\
\includegraphics[width=0.375\textwidth]{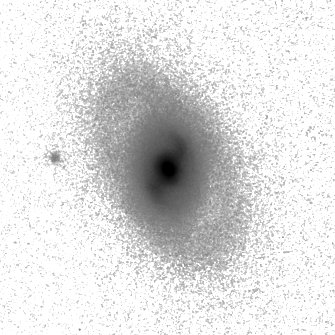}
\includegraphics[width=0.5\textwidth]{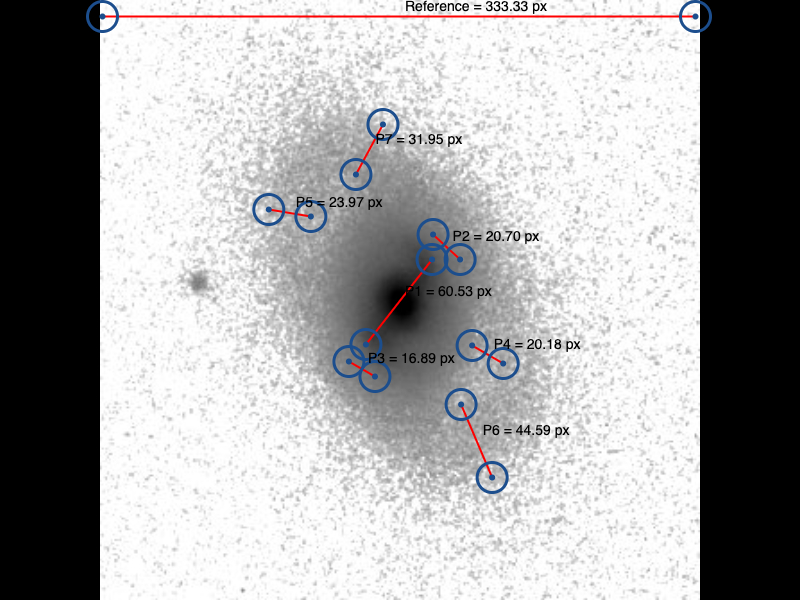}
\caption{The measured sizes of structures in two massive barred galaxies (shown in separate rows). The left-hand column shows the original image of the galaxy, while the right-hand column shows the same image but with measurements of various structures (bar lengths, disk lengths and heights and intra-disk features, such as the width of and spacing between spiral arms) within the galaxy, in units of the number of JWST pixels. Each image is 10 arcseconds on a side, equivalent to $\sim$333 JWST pixels. All images are in the JWST F277W filter. {\color{black}Note that the labels P1, P2, etc. are simply identifiers assigned to individual measurement lines.}} 
\label{fig:bar_detectability}
\end{figure*}


\subsection{Morphological parameters: concentration, asymmetry and clumpiness}
\label{sec:CAS_parameters}

Our study utilizes standard morphological parameters that have been widely used in the literature to quantify the 2D distribution of stellar light in galaxies. Here we briefly define these parameters.

In order to measure our non-parametric CAS morphological parameters, we make use of the \texttt{Morfometryka} code \citep{ferrari2015}, implemented on the JWST F277W images. Using an image stamp of the galaxy alongside the point spread function, the code segments the sources before de-blending them if needed, and masking out any non-central sources. The code then measures the geometry of the galaxy, including the axis ratio, the axis length and the position angle, before defining the Petrosian radius, $R_{\rm petro}$ \citep{Petrosian1976}, where $\langle I(R) \rangle /I(R)=5$ (i.e. the ratio of the mean intensity inside the ellipse and the intensity over the ellipse is equal to 5). A Petrosian region of radius 1.5 $\times$ $R_{\rm petro}$ is then defined, with the same geometric properties as the galaxy, within which all measurements are made. 

The concentration parameter (\textit{C}$_{82}$) is presented in Eq. \ref{eq:c82} below, as defined by \citet{Kent1985}, \citet{Bershady2000} and \citep{Conselice2003}.

\begin{equation}
    \rm \mathit{C}_{82}=5 \times log_{10} \left( \frac{\mathit{R}_{80}}{\mathit{R}_{20}} \right)
    \label{eq:c82}
\end{equation}

\noindent where \textit{R}$_{80}$ and \textit{R}$_{20}$ correspond to the radii enclosing 80 per cent and 20 per cent of the total light of the galaxy. These radii are determined by defining a luminosity growth curve, from which the radius $R_{\rm f}$ containing `f' per cent of the total light is determined.


The asymmetry parameter, as defined by \citet{Abraham1996}, \citet{Conselice2000} and \citet{Rodriguez-Gomez2019}, is described in Eq. \ref{eq:asy} below. 

\begin{equation}
    \rm \mathit{A}=\frac{\sum_{\mathit{i},\mathit{j}} \mid \mathit{I}_{ij} - \mathit{I}_{ij}^{180} \mid }{\sum_{\mathit{i},\mathit{j}} \mid \mathit{I}_{ij} \mid  } - \mathit{A}_{bgr}
    \label{eq:asy}
\end{equation}

\noindent where \textit{I$\rm_{i,j}$} and \textit{I$\rm_{i,j}^{180}$} are the pixel values of the original and the rotated images, respectively. We calculate asymmetry within a circular aperture of 1.5 $\times$ \textit{R}$\rm_{petro}$, with all pixels within this aperture being taken into account for the calculation of asymmetry. \textit{A}$\rm_{bgr}$ is the asymmetry of the background \citep{Lotz2004}. 



The clumpiness parameter (sometimes also referred to as the smoothness index), originally designed by \citet{Conselice2003} and as used in \citet{Lotz2004} and \citet{Rodriguez-Gomez2019}, is described in Eq. \ref{eq:s} below.

\begin{equation}
    \rm \mathit{S}=\frac{\sum_{\mathit{i},\mathit{j}}  \mathit{I}_{ij} - \mathit{I}_{ij}^{S} }{\sum_{\mathit{i},\mathit{j}} \mathit{I}_{ij}} - \mathit{S}_{bgr}
    \label{eq:s}
\end{equation}

\noindent where \textit{I}$\rm_{i,j}$ and \textit{I}$\rm_{i,j}^{S}$ are the pixel values of the original image and its smoothed version, respectively, within circular apertures of 1.5 $\times$ \textit{R}$\rm_{petro}$. The smoothed image is obtained using a boxcar filter of width $\sigma$, which is set to 0.25 $\times$ \textit{R}$\rm_{petro}$, as in \citet[][]{Lotz2004}. The calculation is performed only for the pixels corresponding to radii between $\sigma$ and 1.5 $\times$ \textit{R}$\rm_{petro}$, since the central region is avoided due to most galaxies showing significant central concentration. The pixels with a negative numerator value in Eq. \ref{eq:s} are excluded from the summation. $S\rm_{bgr}$ is the background clumpiness, calculated using the background pixels residing outside the segmentation map.

Errors on the morphological parameters are calculated via a Monte Carlo approach. For each source, and iteration, we inject random Gaussian noise, which is scaled to the measured sky background, $\sigma_{\rm sky}$. From each new image the full analysis pipeline is executed again, including segmentation as well as determining $R_{\rm petro}$ and the centroid of the source, giving us a new measure of the parameters of interest. The uncertainties are estimated from 500 repetitions of this process. The median errors in the concentration, asymmetry and clumpiness parameters are 2.1, 9.6 and 27.8 per cent respectively. 


{\color{black}We have verified that the distributions of CAS parameters for massive galaxies remain unchanged in the redshift range $z<0.4$, with the median values for galaxies at $z < 0.2$ being virtually identical to those for their counterparts at $0.2<z<0.4$.} This indicates that using a redshift range for the massive galaxy sample that is slightly larger than that of our dwarf population does not alter our results.

\subsection{Morphological classification and identification of barred dwarf galaxies via visual inspection}
\label{sec:morphological_classification}

We visually inspect JWST F277W and HSC $gri$ colour composite images of each galaxy to both morphologically classify the system and identify galaxies that exhibit bars. Since structures within dwarf galaxies may exhibit lower contrast than those in their massive counterparts \citep{Lazar2024a}, we also inspect `unsharp masked' versions of the JWST and HSC images. Unsharp masking \citep[e.g.][]{Malin1977} results in the sharpening of the edges of structures, like bars, contained within the object. It has previously been used in astrophysical image analysis to detect faint, low-contrast features like shells and tidal features inside and around nearby massive galaxies \citep[see e.g.][]{Malin1983}. 

The visual inspection is carried out by one expert classifier (SK). The images are randomised, with both the original and unsharp-masked images classified at the same time. Physical parameters such as the stellar mass and redshift are kept hidden to avoid introducing biases into the classification process. 


\subsubsection{Morphological classification into early-type and late-type galaxies}
\label{sec:etg_ltg_classification}

We use our visual inspection to separate the galaxy population into three broad morphological classes. The first are `early-type' galaxies (ETGs) which have central light concentrations but otherwise exhibit smooth light distributions. The second are `late-type' galaxies (LTGs) which lack a central light concentration that is typical of ETGs and often show structure (e.g. spiral arms, clumps etc.). The final category, which only appears in the dwarf regime, consists of `compact' objects which, while resolved in the JWST images, are somewhat too small to classify securely. A very small number of objects cannot be classified, either because the objects do not have enough flux in the JWST image to make classification possible or because they appear to be a blend of two objects in the ground-based HSC image (which could compromise the COSMOS2020 photometry). Around 98 per cent of our dwarf galaxies are classifiable. {\color{black}Table \ref{tab:fetg} presents the fraction of ETGs in three different mass ranges: the lower (10$^7$ M$_{\odot}$ < $M_{\star}$ < 10$^{8.25}$ M$_{\odot}$) and upper (10$^{8.25}$ M$_{\odot}$ < $M_{\star}$ < 10$^{9.5}$ M$_{\odot}$) halves of the stellar mass range in the dwarf regime and the massive-galaxy regime ($M_{\star}$ > 10$^{10}$ M$_{\odot}$). The uncertainties shown are from counting statistics only. We note, however, that, given the small area of COSMOS and the relatively narrow redshift range of our study, quantities like the ETG fraction are subject to a large uncertainty due to cosmic variance. For example, the fractional error due to cosmic variance on the number counts of $\sim$1000 galaxies in such a region is expected to be greater than 25 per cent \citep[e.g.][]{Moster2010}. {\color{black}While the uncertainties} in Table \ref{tab:fetg} show only those due to counting statistics, it is worth bearing in mind that the uncertainty incurred due to cosmic variance is significantly higher.} Figure \ref{fig:example_galaxy_images} presents examples of galaxies in our different morphological categories. 

\begin{table}
\begin{center}
\begin{tabular}{ l | l | l | l }
\toprule
Stellar mass range & $f_{\rm ETG}$ & $f_{\rm LTG}$ & $f_{\rm compact}$\\
\midrule
7.00 < Log (M$_{\star}$/M$_{\odot}$) < 8.25 & 0.351$^{\pm 0.017}$ & 0.598$^{\pm 0.017}$ & 0.028$^{\pm 0.006}$\\
8.25 < Log (M$_{\star}$/M$_{\odot}$) < 9.50 & 0.505$^{\pm 0.033}$ & 0.486$^{\pm 0.033}$ & 0.005$^{\pm 0.006}$\\
Log (M$_{\star}$/M$_{\odot}$) > 10         & 0.522$^{\pm 0.012}$   & 0.478$^{\pm 0.012}$ & 0\\
\bottomrule
\end{tabular}
\caption{{\color{black}The fraction of ETGs in three different mass ranges. The uncertainties shown are from counting statistics only. Given the small area of COSMOS and the relatively narrow redshift range of our study, quantities like the ETG fraction are subject to a large uncertainty due to cosmic variance. This is likely to be significantly larger than that due to the counting statistics alone. See text in Section \ref{sec:etg_ltg_classification} for details.}} 
\label{tab:fetg}
\end{center}
\end{table}

\begin{figure}
\includegraphics[width=\columnwidth]{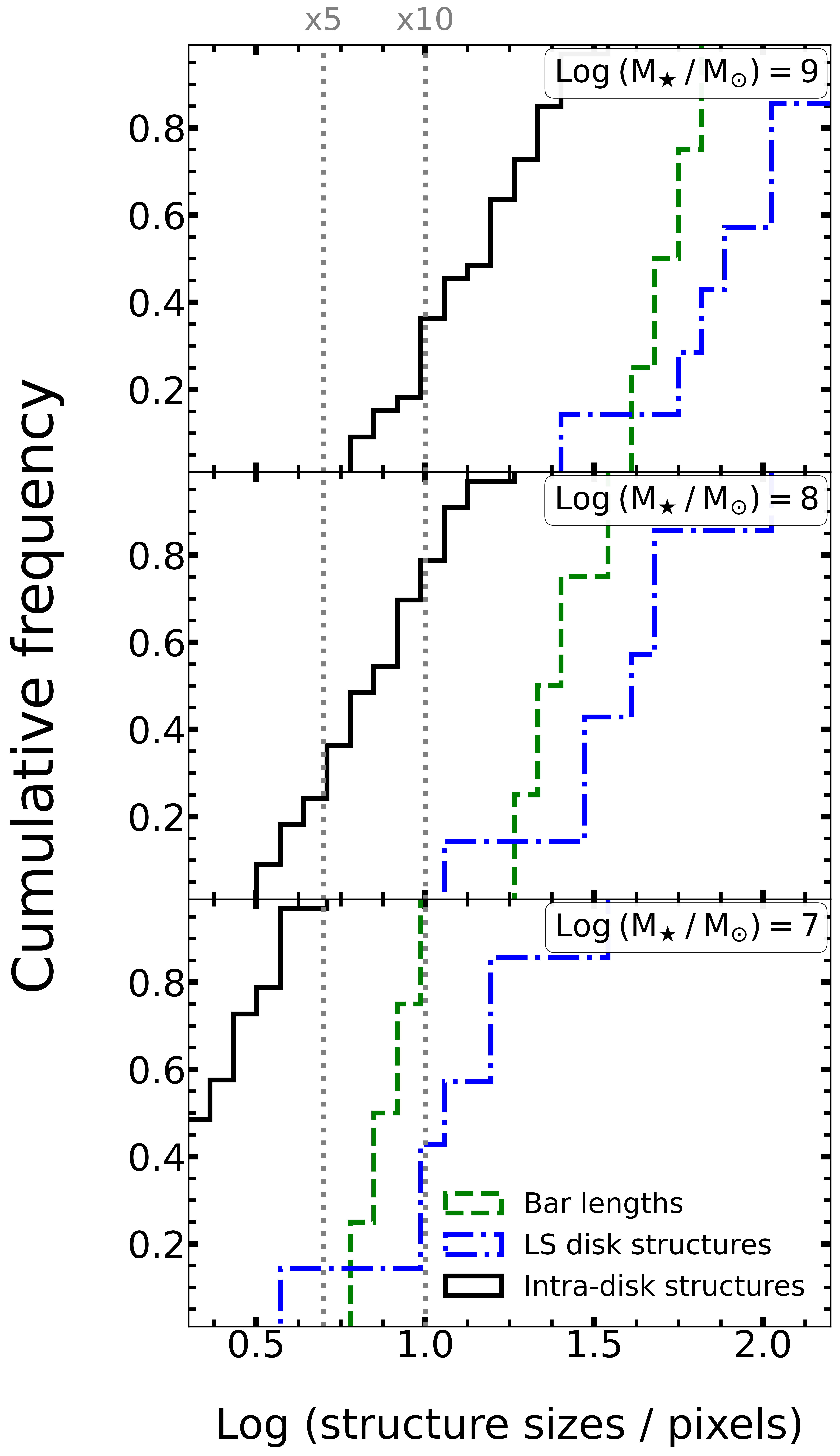}
\caption{Cumulative distributions of the predicted sizes, in JWST pixels, of bar lengths, large-scale (LS) disk structures (such as disk lengths and heights) and intra-disk features in dwarf galaxies at various stellar masses. The top, middle and bottom panels show predictions for dwarf galaxies with $M_{\star}$ $\sim$ 10$^9$, 10$^8$ and 10$^7$ M$_{\odot}$ respectively. The dotted vertical lines indicate 5 and 10 JWST pixels.}
\label{fig:structure_measurements}
\end{figure}

\begin{figure*}
\includegraphics[width=0.9\textwidth]{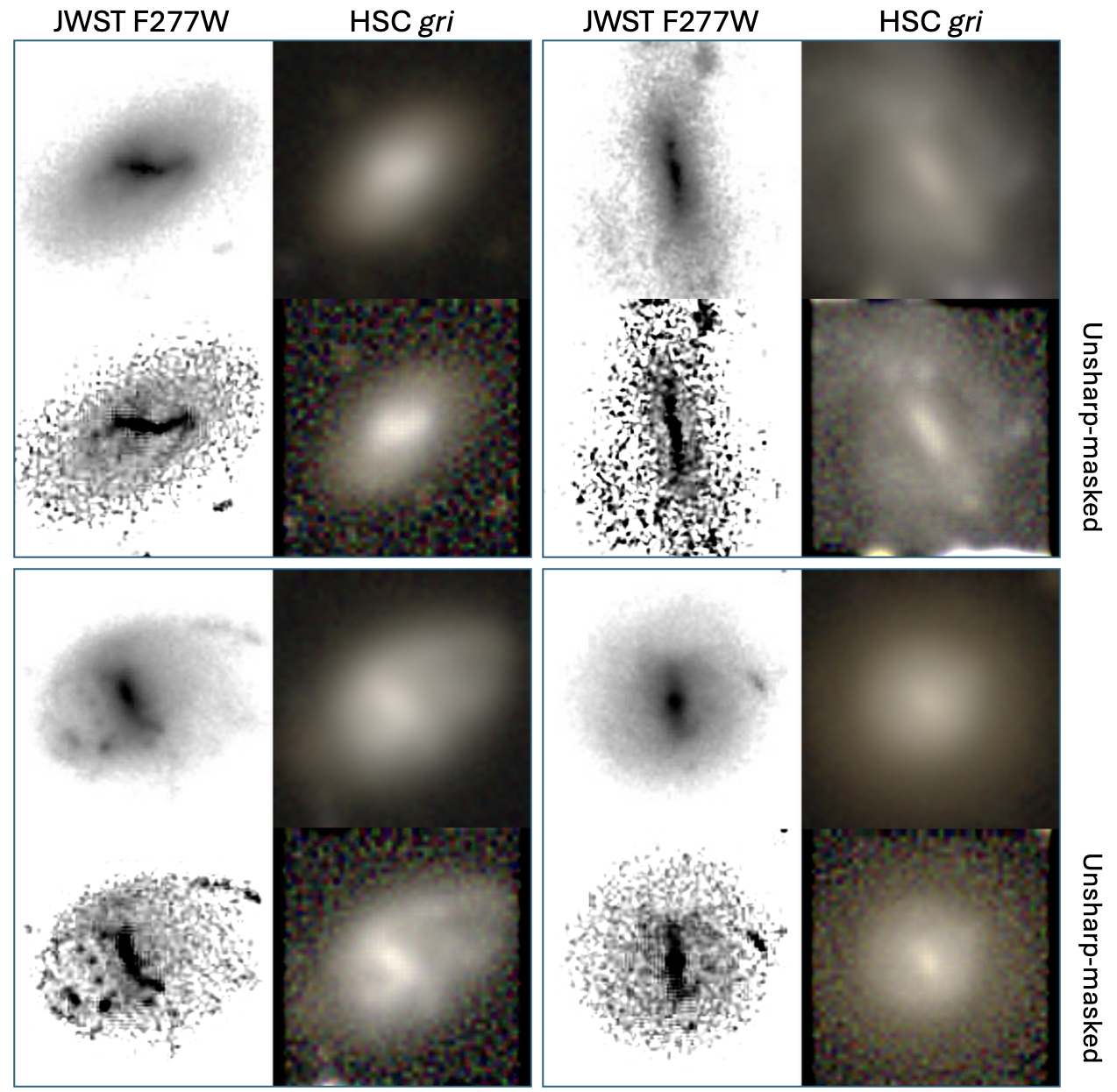}
\caption{{\color{black}Examples of dwarf galaxies in our sample which have bars, identified via visual inspection. The size of each image is 5 arcseconds on a side. The images from different instruments have different orientations - we intentionally keep it this way so that each galaxy is visually inspected at different orientations. Each image has four panels. The first and second columns show the JWST F277W and HSC ($gri$) colour composite images respectively. The first row shows the original images, while the second row shows their unsharp-masked counterparts.}}
\label{fig:example_barred_galaxy_images}
\end{figure*}


\subsubsection{Identification of barred dwarf galaxies}
\label{ref:bar_identification}

We use our visual inspection to identify galaxies which contain bars. It is worth considering first whether bars are likely to be detectable in dwarfs at various stellar masses, given the depth and resolution of the JWST images available to us. To do this, we first measure the sizes of bars that are clearly present in the JWST images of massive galaxies in our redshift range of interest ($z<0.15$). We then use the stellar mass -- galaxy size relation in the nearby Universe, measured via the DESI Early Data Release \citep{DESI2024}, to predict the sizes that bars are likely to have, in dwarfs of various stellar masses, if they did indeed exist in these systems. Comparing these predicted sizes to the resolution of JWST then enables us to estimate a lower stellar mass limit down to which bar identification via visual inspection is likely to succeed using the JWST images.

Figure \ref{fig:bar_detectability} shows example JWST images of two massive galaxies which have bars (both of which have $M_{\star}$ > 10$^{10}$ M$_{\odot}$). The left-hand column shows the original image of the galaxy, while the right-hand column shows the same image but with measurements of various structures within the galaxy, in units of the number of JWST pixels\footnote{{\color{black}The lengths of the structures in Figure \ref{fig:bar_detectability} have been measured using the following online tool: https://imagemeasurementonline.com/}}. For the purposes of this exercise, the structures we consider are bar lengths, disk lengths and heights and intra-disk features such as the width of and spacing between spiral arms. While our analysis is simply focused on bars, it is useful to consider other types of structures because it helps us understand how the dimensions of the bars that we are eventually able to study compares to the spatial scales of various structures in these galaxies. 

Once the aforementioned structures are measured in the JWST images of massive galaxies at $z<0.15$, we use the stellar mass -- galaxy size relation, described above, to predict the sizes that they would have in galaxies of different stellar masses. Figure \ref{fig:structure_measurements} summarizes the cumulative distributions of the predicted  bar lengths (red), disk lengths and heights (blue) and sizes of intra-disk features (black), in galaxies with stellar masses of 10$^9$ M$_{\odot}$ (top row), 10$^8$ M$_{\odot}$ (middle row) and 10$^7$ M$_{\odot}$ (bottom row) respectively. The grey dashed vertical lines indicate multiples of JWST pixels. It is apparent that at 10$^9$ M$_{\odot}$ (10$^8$ M$_{\odot}$), the predicted bar sizes are at least 40 (20) JWST pixels wide and should therefore be easily identifiable. However, at 10$^7$ M$_{\odot}$, the majority of bars (if they did exist) are likely to be less than 10 pixels in length, suggesting that it will be challenging to detect them clearly, even with the resolution of JWST. 

Note also that the lengths of the detectable bars are likely to be similar to large-scale disk structures (e.g. disk scale lengths and heights) so our study is primarily sensitive to `strong' rather than `weak' bars. Since our analysis above indicates that we are likely to be able to quantify the bar fraction down to 10$^8$ M$_{\odot}$ at $z<0.15$, we restrict our exploration of the bar fraction, in Section \ref{sec:bar_fraction} below, to galaxies with $M_{\star}$ > 10$^8$ M$_{\odot}$. Figure \ref{fig:example_barred_galaxy_images} presents examples of barred dwarf galaxies in our sample that are identified via our visual inspection process.


\section{Results}

\subsection{Trends in morphology with stellar mass and effective surface brightness}

\label{sec:morphology_mass}

\begin{figure}
\includegraphics[width=0.95\columnwidth]{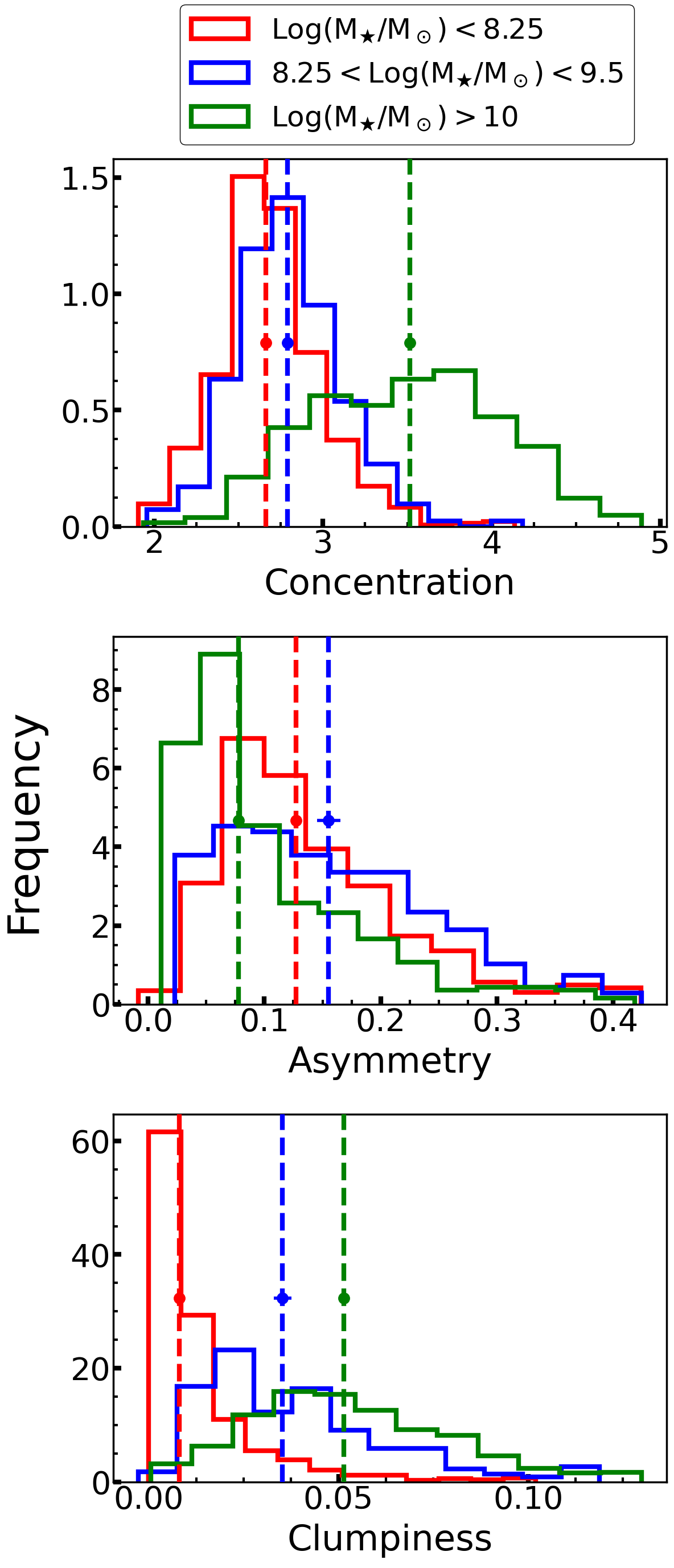}
\caption{The concentration (top), asymmetry (middle) and clumpiness (bottom) distributions for galaxies in three different mass regimes: the lower (10$^7$ M$_{\odot}$ < $M_{\star}$ < 10$^{8.25}$ M$_{\odot}$; red) and upper (10$^{8.25}$ M$_{\odot}$ < $M_{\star}$ < 10$^{9.5}$ M$_{\odot}$; blue) halves of the stellar mass range in the dwarf regime and the massive galaxy regime ($M_{\star}$ > 10$^{10}$ M$_{\odot}$; green). For each distribution we show the median as a vertical dashed line with an associated error bar calculated via standard bootstrapping.}
\label{fig:CAS_mass}
\end{figure}

\begin{figure*}
\includegraphics[width=0.95\textwidth]{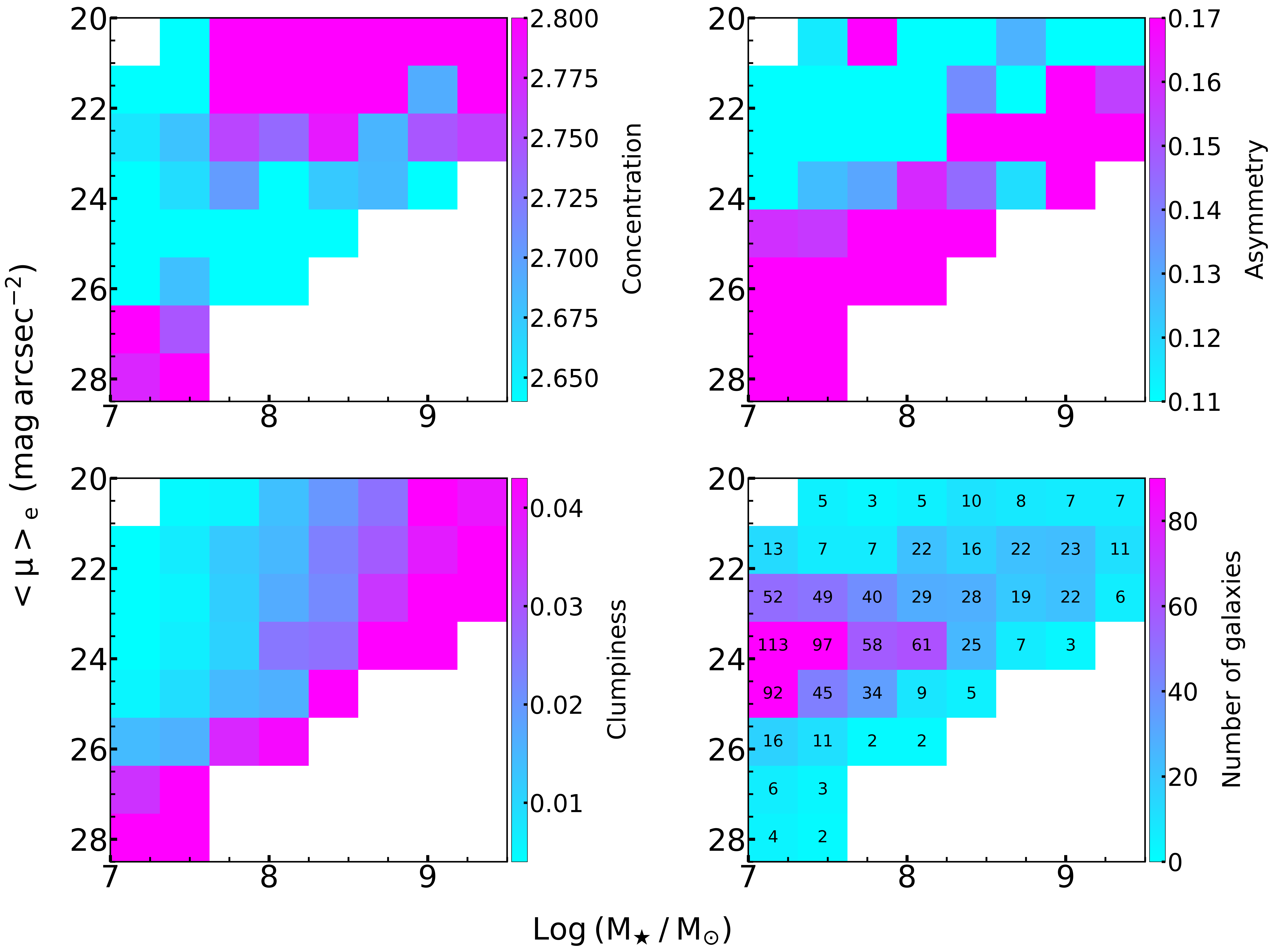}
\caption{The concentration (top-left), asymmetry (top-right) and clumpiness (bottom-left) distributions in the effective surface brightness -- stellar mass plane. The bottom-right panel shows the number of galaxies in different regions of this plane.}
\label{fig:CAS_effSB_mass}
\end{figure*}

\begin{table}
\begin{center}
\begin{tabular}{ c | c }
\toprule
& Rank correlation coefficient\\
\midrule
C vs Log M$_{\star}$ & +0.67\\
A vs Log M$_{\star}$ & -0.24\\
S vs Log M$_{\star}$ & +0.75
\end{tabular}
\caption{Results of Spearman's rank correlation tests between the CAS parameters and stellar mass.} 
\label{tab:spearman}
\end{center}
\end{table}


Figure \ref{fig:CAS_mass} shows the distributions of CAS parameters in three different mass regimes: the lower (10$^7$ M$_{\odot}$ < M$_{\star}$ < 10$^{8.25}$ M$_{\odot}$) and upper (10$^{8.25}$ M$_{\odot}$ < M$_{\star}$ < 10$^{9.5}$ M$_{\odot}$) halves of the stellar mass range in the dwarf regime and the massive galaxy regime (M$_{\star}$ > 10$^{10}$ M$_{\odot}$). The median values of the distributions (indicated using the dashed vertical lines), indicate that all parameters exhibit systematic trends with stellar mass. Both concentration and clumpiness decline at lower stellar masses (i.e. as we move from the massive to the dwarf-galaxy regime), {\color{black}while asymmetry shows an opposite (albeit weaker) trend} and increases with decreasing stellar mass. To further quantify these trends, we perform Spearman rank correlation tests between each morphological parameter and stellar mass, across the massive and dwarf-galaxy regimes. This test returns a correlation coefficient between +1 and -1, where a positive (negative) value indicates a correlation (anti-correlation) between the variables. The result can be considered significant if the associated p-value is less than the standard threshold of 0.05. 

Table \ref{tab:spearman} presents the correlation coefficients between the CAS parameters and stellar mass across the dwarf and massive-galaxy regimes, which confirm the visual trends that are apparent in Figure \ref{fig:CAS_mass}. Note that all tests return very low p-values ($p \ll 0.05$), indicating that the trends between the parameters are significant. Both concentration and clumpiness show a strong positive correlation with stellar mass, while asymmetry shows a moderate negative correlation with stellar mass. In this context, recall from the introduction that past surveys like the SDSS, which are shallow, only enable access to the bluest dwarf galaxies outside the very local Universe. In Appendix \ref{app:bluest_dwarfs} we demonstrate how our results change if we restrict the dwarf population to the bluest ten percentiles of the rest-frame colour distribution. Figure \ref{fig:CAS_mass_bluest_dwarfs} demonstrates that the bluest dwarfs have higher asymmetries and clumpiness than a mass-complete sample, which dilutes the quantitative trends (see Table \ref{tab:blue_spearman}) presented above. 

It is instructive to explore whether trends may exist between morphology and the overall distribution of baryons within individual galaxies that have similar stellar masses. Figure \ref{fig:CAS_effSB_mass} shows the values of the CAS parameters as a heatmap, as a function of both the stellar mass and the effective surface brightness ($\langle$$\mu$$\rangle$$_{\rm e}$) of galaxies. Note that $\langle$$\mu$$\rangle$$_{\rm e}$ is defined here as the average surface brightness within the half-light radius of the galaxy. Figure \ref{fig:CAS_effSB_mass} indicates that there are no strong trends in concentration as a function of $\langle$$\mu$$\rangle$$_{\rm e}$ at fixed stellar mass in the dwarf regime. Interestingly, however, there is a hint that, while concentration typically has a tendency of decreasing as $\langle$$\mu$$\rangle$$_{\rm e}$ becomes fainter at a given stellar mass, this trend appears to reverse at the lowest stellar masses considered here. 

While asymmetry appears to increase for objects with fainter $\langle$$\mu$$\rangle$$_{\rm e}$ at a fixed stellar mass, {\color{black}clumpiness shows only a weak trend with $\langle$$\mu$$\rangle$$_{\rm e}$}. The bottom-right panel of Figure \ref{fig:CAS_effSB_mass} shows the numbers of galaxies in the effective surface brightness -- stellar mass plane, which peak at M$_\star$ $\sim$ 10$^{7.2}$ M$_\odot$ and $\langle$$\mu$$\rangle$$_{\rm e}$ $\sim$ 24 mag arcsec$^{-2}$. In Appendix \ref{app:CAS_errors} we show, by perturbing the CAS values using their errors, that the structure of these heatmaps do not change. In other words, the trends seen in this plane are not washed out by the uncertainties in the CAS parameters. While our focus is on dwarf galaxies here, we extend this plot to the massive galaxy regime in Appendix \ref{app:CAS_eff_massive}. 


The systematic decrease in concentration observed at progressively lower stellar mass is likely to be driven by two principal reasons. First, shallower potential wells are less able to retain gas in the central regions of the galaxy, as a result of which star formation is likely to take place in a spatially distributed manner rather than in a compact central region. Second, baryonic feedback (e.g from supernovae or AGN) is likely to be more effective at stirring up gas and moving it away from the central regions, flattening the central gas density profiles. These flatter profiles will then be inherited by the stars that form from that gas \citep[e.g.][]{Jackson2021b}. The observed increase in concentration at the faintest values of $\langle$$\mu$$\rangle$$_{\rm e}$ at the very lowest stellar masses considered here (top left panel of Figure \ref{fig:CAS_effSB_mass}) may be caused by the impact of these processes becoming extreme. For example, the gas outflows may become large enough to disrupt any stellar assembly beyond the initial stellar mass buildup in the centre, which then produces a faint system with relatively high concentration. Furthermore, the removal of a significant amount of gas from the potential well will prevent subsequent activity (e.g. supernova feedback) which could make the mass profile in the central regions flatter \citep[e.g.][]{Koudmani2025}.

The anti-correlation between asymmetry and stellar mass, which mirrors the findings of \citet{Lazar2024a}, is likely driven by the fact that shallower potential wells are more sensitive to any process that acts to alter the distribution of gas and stars in the galaxy e.g. tidal perturbations, baryonic feedback etc. Finally, the trend of decreasing clumpiness with lower stellar mass is likely to have two principal causes. First, dwarf galaxies exhibit lower gas surface densities than their massive counterparts \citep[e.g.][]{Leroy2005,Bigiel2008,Roychowdhury2015,Pessa2021}. Second, there is a gradual change in galaxy kinematics as stellar mass decreases. The gas content of dwarf late-types becomes increasingly more dispersion-dominated than in their massive counterparts \citep[e.g.][]{Moiseev2012,Ianjamasimanana2015,Lelli2022}, which is also reflected in their stellar kinematics \citep[e.g.][]{Wheeler2017,Falconbarroso2019,Scott2020}. The change in the clumpiness can then be interpreted in the context of the Toomre parameter \citep[$Q$;][]{Toomre1964}. Both an increase in velocity dispersion and a decrease in the gas surface density would increase the value of $Q$, which leads to weaker fragmentation and a lower value of clumpiness \citep[see e.g.][]{Chantavat2024}. Although the clumpiness parameter in our study relates to the stellar light distribution, it reflects the spatial distribution of young stellar populations and therefore the fragmentation of the underlying gas disk. 



\subsection{Trends between morphology and recent star formation history as a function of stellar mass}
\label{sec:CAS_recent_sfh}

We proceed by exploring how the relationship between morphology and the recent SFH may change as a function of stellar mass. We study the recent SFH via the fraction of galaxies that are red (i.e. quenched of star formation). Following \cite{Kaviraj2025} the threshold rest-frame $(g-i)$ values that separate red and blue galaxies are taken to be 0.7 and 0.9 in the dwarf and massive-galaxy regimes respectively.

\begin{figure}
\includegraphics[width=0.95\columnwidth]{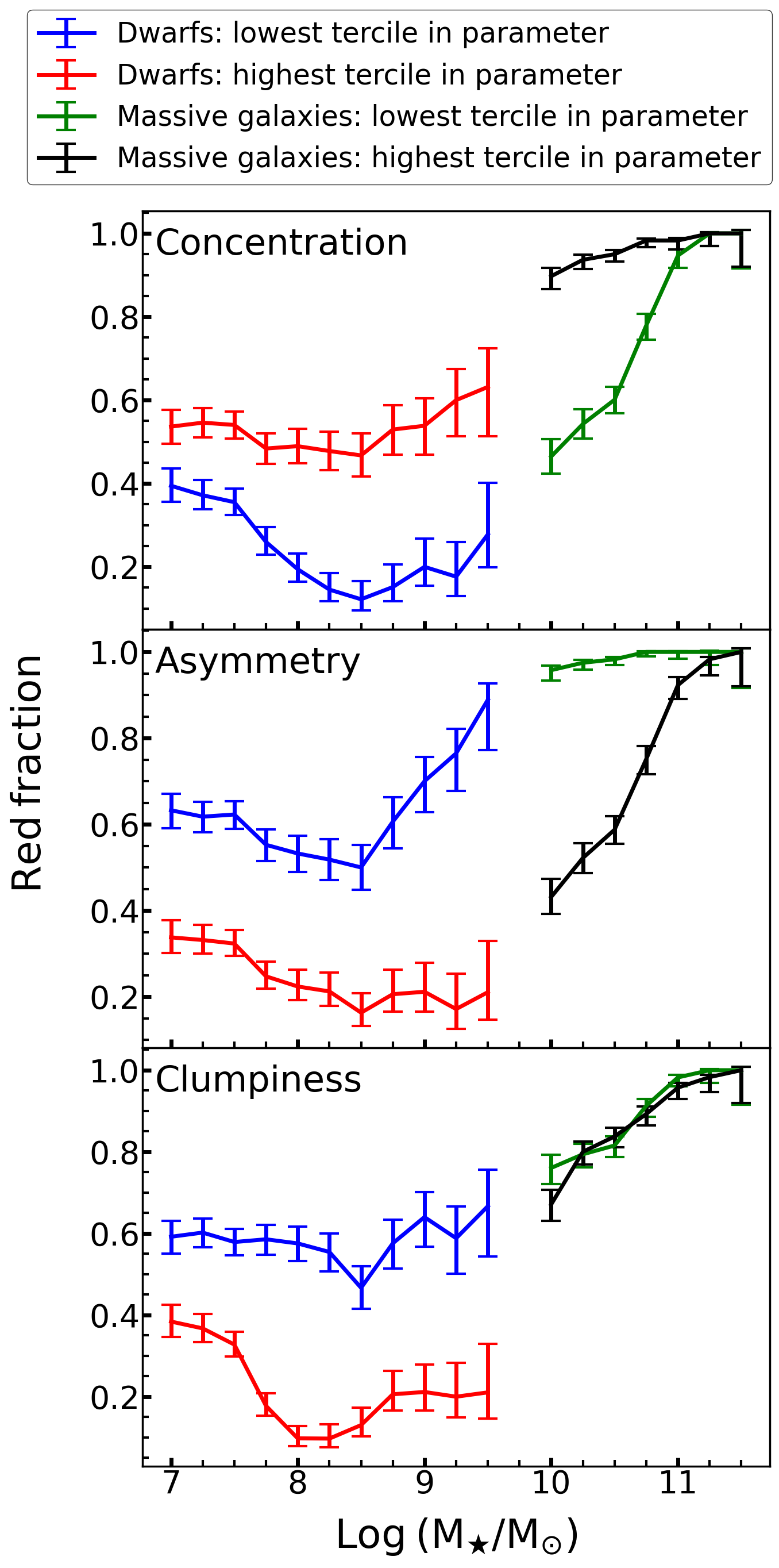}
\caption{The red fraction for galaxies in the lowest and highest terciles in concentration (top), asymmetry (middle) and clumpiness (bottom). The lowest and highest terciles are shown in blue and red for dwarfs and green and black for massive galaxies.}
\label{fig:CAS_red_fraction}
\end{figure}

For this exercise, we split our massive and dwarf-galaxy populations into three terciles in the concentration, asymmetry and clumpiness parameters, at fixed stellar mass. We then study the galaxy red fraction for objects in the upper, middle and lower terciles in each parameter, as a function of stellar mass. Figure \ref{fig:CAS_red_fraction} indicates that, at a given stellar mass, the red fraction is highest and lowest for galaxies in the highest and lowest terciles in concentration respectively. In other words, higher concentration is systematically correlated with a greater probability for the galaxy to be quenched, largely irrespective of stellar mass. 

The correlation between concentration and redness can be explained by the fact that more concentrated objects typically have more pronounced bulges. This can lead to more effective quenching as the black hole mass, which correlates with the effectiveness of AGN feedback \citep[e.g.][]{Fabian2012}, tends to correlate with the mass of the bulge \citep[e.g.][]{Haring2004,Martin-navarro2018,Ferre-mateu2021,Bluck2014} and also because morphological quenching is expected to be enhanced in systems with more prominent bulges \citep[e.g.][]{Martig2009,Gensior2020}. 

In a similar vein, the red fraction is systematically higher at progressively smaller values of asymmetry because star formation correlates with asymmetry in the stellar body of the galaxy \citep[e.g.][]{Abraham1996,Conselice2003}. A similar trend is seen between the red fraction and clumpiness in the dwarf regime, although this effect appears to be absent in massive galaxies. This can be explained by the fact that the increase in the velocity dispersion and decrease in the gas surface density in lower mass galaxies both act to reduce the SFR which then drives a decrease in the clumpiness. {\color{black}The lack of a trend at very high stellar mass is likely driven by the fact that, in this regime, virtually all galaxies are red, devoid of SFR (plausibly due to internal processes, e.g. \citealt{Peng2010,Bluck2020}) and have a narrow range of clumpiness values.} 

\subsection{ETGs and LTGs become more difficult to separate parametrically as stellar mass decreases}
\label{sec:CAS_to_morphology}

The progressively lower values of concentration with decreasing stellar mass have an important consequence for our ability to separate dwarf ETGs and LTGs using morphological parameters. This has already been noted by \citet{Lazar2024a}, who have demonstrated, using ground-based HSC images (only), that it becomes challenging to separate the different morphological classes that exist in the dwarf regime using commonly-used parameters like CAS (as well as others like the Sérsic index, $M_{\rm 20}$, and the Gini coefficient). This is caused by the fact that much of the parametric discrimination between ETGs and LTGs in the massive galaxy regime is driven by the fact that massive ETGs are significantly more concentrated than massive LTGs. However, as both ETGs and LTGs become less concentrated in the dwarf regime, the discriminatory leverage offered by this parameter decreases, making it difficult to separate morphological classes via parameters in the dwarf regime. 

\begin{figure}
\includegraphics[width=1\columnwidth]{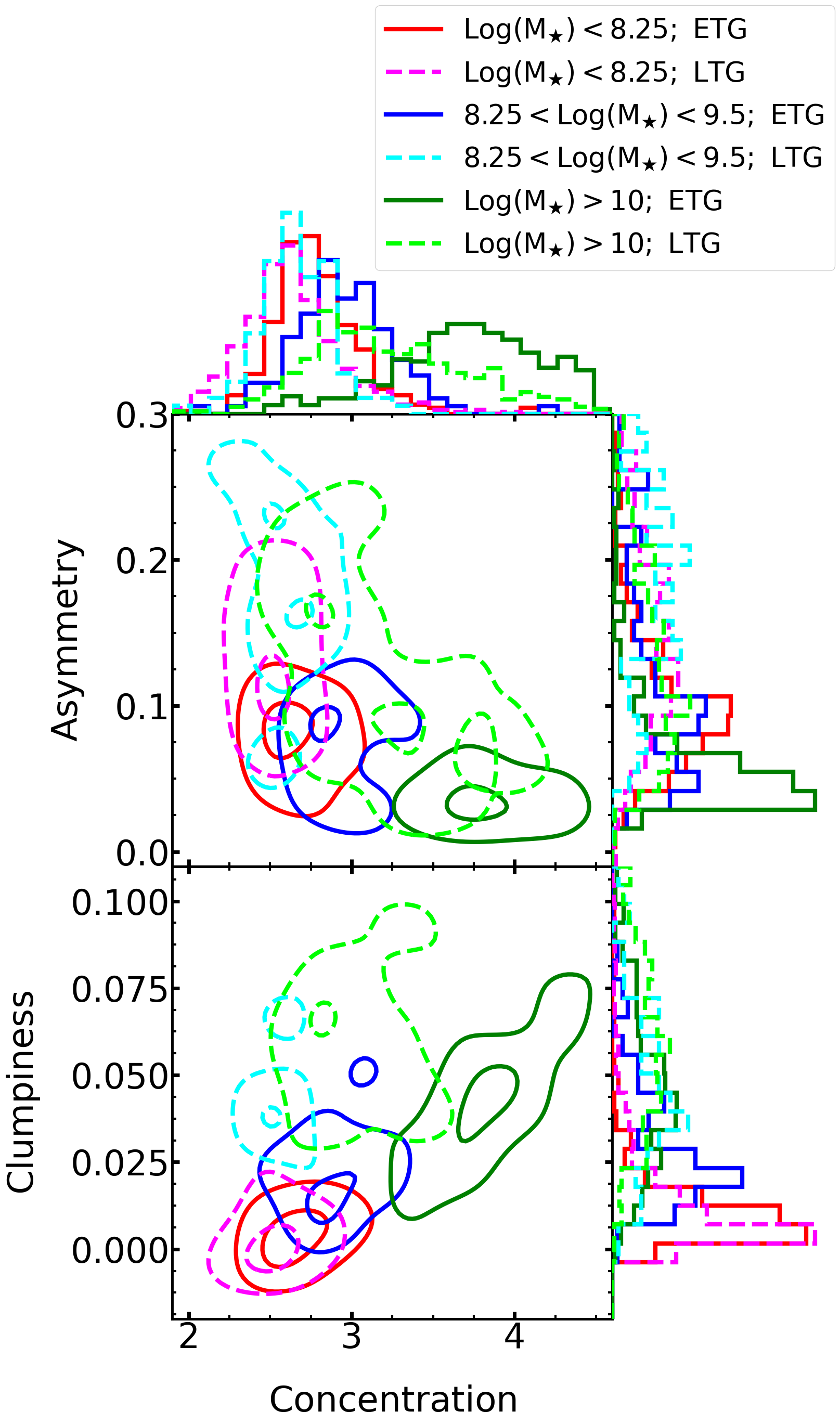}\\
\caption{The asymmetry vs concentration (top) and clumpiness vs concentration planes (bottom) for galaxies in three different mass regimes: the lower (10$^7$ M$_{\odot}$ < $M_{\star}$ < 10$^{8.25}$ M$_{\odot}$) and upper (10$^{8.25}$ M$_{\odot}$ < $M_{\star}$ < 10$^{9.5}$ M$_{\odot}$) halves of the stellar mass range in the dwarf regime and the massive galaxy regime ($M_{\star}$ > 10$^{10}$ M$_{\odot}$).}
\label{fig:CAS_visual}
\end{figure}

Here, we revisit this analysis at JWST resolution (which has a linear resolution that is twenty times better than that of HSC), down to $M_{\star}$ $\sim$ 10$^{7}$ M$_{\odot}$ (while the \citet{Lazar2024a} analysis was restricted to $M_{\star}$ > 10$^{8}$ M$_{\odot}$). Figure \ref{fig:CAS_visual} presents the asymmetry vs concentration and clumpiness vs concentration planes in three different mass regimes: the lower (10$^7$ M$_{\odot}$ < $M_{\star}$ < 10$^{8.25}$ M$_{\odot}$) and upper (10$^{8.25}$ M$_{\odot}$ < $M_{\star}$ < 10$^{9.5}$ M$_{\odot}$) halves of the stellar mass range in the dwarf regime and the massive galaxy regime ($M_{\star}$ > 10$^{10}$ M$_{\odot}$). In a similar vein to the conclusions of \citet{Lazar2024a}, we find that the separation between ETGs and LTGs, in both parameter planes, steadily decreases as we move towards lower stellar masses. The positions of ETGs and LTGs in these planes indicate that they can be separated, via parameters, in the massive-galaxy regime and, to a reasonable extent, in the upper half of the stellar mass range in the dwarf regime. However, in the lower half of the dwarf stellar mass range, ETGs and LTGs lie very close to each other in either plane, making it hard to separate them parametrically.

\subsection{Bars become less frequent with decreasing stellar mass}
\label{sec:bar_fraction}

We complete our study by examining the strong bar fraction in our dwarf sample, defined as the fraction of galaxies that are not edge-on but have visually-identified strong bars {\color{black}(examples of such edge-on galaxies are shown in Appendix \ref{app:edge_on_galaxies})}. Recall from the detectability analysis in Section \ref{ref:bar_identification} above that strong bars are likely to be identifiable down to $M_{\star}$ $\sim$ 10$^8$ M$_{\odot}$. Table \ref{tab:bar_fractions} presents bar fractions within the dwarf regime (10$^8$ M$_{\odot}$ < M$_{\star}$ < 10$^{9.5}$ M$_{\odot}$) from our study, as well as values of the bar fraction in the massive regime from the recent literature. Together with the literature, our results indicate that the bar fraction decreases steadily towards lower stellar masses and is consistent with zero around 10$^8$ M$_{\odot}$, suggesting a lower limit to the galaxy stellar mass that is required to induce bar formation. 

While bars have been extensively studied in massive galaxies \citep[e.g.][]{Masters2011}, explorations of the bar fraction in relatively unbiased, statistically significant populations of dwarfs remain sparse, with previous efforts largely focusing on local ($z<0.02$) galaxies in relatively dense environments. For example, \citet[][]{Mendez-Abreu2010}, who employ visual inspection of HST images, find no barred dwarfs in a sample of $\sim$190 galaxies with M$_{\star}$ < 10$^{9}$ M$_{\odot}$ in the Coma cluster. {\color{black}\citet{Mendez-Abreu2012} show that the decrease in bar fraction at lower stellar masses takes place regardless of the local environment.} Similarly, \citet[][]{Marinova2012} use visual inspection, coupled with unsharp masking, of the HST images of 333 dwarf galaxies (M$_{\star}$ < 10$^{9.5}$ M$_{\odot}$) in Coma to arrive at a similar conclusion, with only 4 per cent of their sample exhibiting bars. The results of both studies appear consistent with our conclusions. While these studies focus on dense environments, there is some evidence that the incidence of bars is not strongly correlated with local environments \citep[e.g.][]{Martinez2011,Marinova2012,Lee2012}, making these studies comparable with our work. {\color{black}It is worth noting that \citet{Mendez-Abreu2012} attribute the decrease in the bar fraction at progressively lower stellar masses to an increase in the disk thickness, with the disks of lower stellar mass galaxies being more fragile and hot which, in turn, inhibits the formation of bars.}

We note, however, that some recent studies do come to different conclusions, although these discrepancies may be driven by selection effects. For example, \citet{Nair2010} study barred galaxies in the SDSS at $z<0.1$ and find that the barred fraction increases in their sample towards lower stellar masses, reaching $\sim$40 per cent at M$_{\star}$ $\sim$ 10$^{9}$ M$_{\odot}$. However, this result is likely to be strongly influenced by the selection bias described in Section \ref{sec:C2020_sample}, since low-mass galaxies in this sample are dominated by blue (i.e. gas rich), late-type systems (see their Figure 1) in which bars are more likely to form. \citet{Geron2021} perform a similar analysis using the SDSS, but at lower redshifts $(z<0.05$), and find a lower bar fraction ($\sim$10 per cent at M$_{\star}$ $\sim$ 10$^9$ M$_{\odot}$). A lower bar fraction in a galaxy sample that is restricted to lower redshift is consistent with the hypothesis that SDSS bar fractions are affected by strong selection biases in the dwarf regime. Not unexpectedly, restricting the galaxy population to lower redshifts reduced the bias towards late-type systems and moves the bar fraction closer to the values reported in our work.  

\citet[][]{Erwin2018}, who use a sample of $\sim$600 galaxies within 40 Mpc with 10$^8$ M$_{\odot}$ < M$_{\star}$ < 10$^{11.5}$ M$_{\odot}$ from the S$^{4}$G survey report a decline in the bar fraction as galaxy stellar mass decreases, as is the case in this study. 
However, their dwarf bar fractions themselves (e.g. $\sim$40 per cent at M$_{\star}$ $\sim$ 10$^{8.5}$ M$_{\odot}$) are larger than those found in our study. This disagreement is likely to be driven by the fact that the Spitzer data used by S$^{4}$G is $\sim$5 mag shallower than our JWST images and because S$^{4}$G imposes an angular limit of greater than 1 arcmin as part of its selection criteria. This is likely to skew the dwarf sample within this survey towards spatially extended late-type systems, making a direct comparison to our results challenging.

\begin{table}
\begin{center}
\begin{tabular}{ c | c | c}
\toprule
Stellar mass range & (Strong) bar fraction  & Source\\
\midrule
8.0 < Log (M$_{\star}$/M$_{\odot}$) < 8.5 & 0.01$^{\pm 0.01}$ & This work\\
8.5 < Log (M$_{\star}$/M$_{\odot}$) < 9.0 & 0.05$^{\pm 0.02}$  & This work\\
9.0 < Log (M$_{\star}$/M$_{\odot}$) < 9.5 & 0.09$^{\pm 0.04}$  & This work \\
 Log (M$_{\star}$/M$_{\odot}$) $\sim$ 10 & 0.16 & \citet{Geron2021}\\
  Log (M$_{\star}$/M$_{\odot}$) $\sim$ 10.6 & 0.23 & \citet{Geron2021}\\
\bottomrule
\end{tabular}
\caption{{\color{black}Bar fractions, defined as the fraction of LTGs that are not edge-on but have visually-identified bars, in various stellar mass ranges in the dwarf regime. Uncertainties are calculated following \citet{Cameron2011} and are shown using superscripts.}} 
\label{tab:bar_fractions}
\end{center}
\end{table}


\section{Summary}
\label{sec:summary}

We have used a mass-complete sample of $\sim$1000 galaxies at $z<0.15$, to explore how the trends between galaxy morphology and other physical parameters evolve as a function of stellar mass down to $M_{\star}$ $\sim$ 10$^{7}$ M$_{\odot}$. Our study combines morphological parameters (concentration, asymmetry and clumpiness) with visual morphological classifications to explore: (1) how galaxy morphology evolves with stellar mass and effective brightness, (2) the connection between morphology and recent star formation history as a function of stellar mass, (3) the incidence of bars in the dwarf regime and (4) how well morphological parameters perform in separating early and late-type galaxies {\color{black}as a function of stellar mass}. Our main conclusions are as follows:

\begin{itemize}

    \item Galaxies become progressively less concentrated, more asymmetric and less clumpy at lower stellar masses. The decrease in concentration and increase in asymmetry are likely to be driven by the fact that shallower potential wells are less able to hold material in their central regions and are also more susceptible to processes (e.g. baryonic feedback and tidal perturbations) which can alter the distribution of gas and stars in the galaxy. The decrease in clumpiness is likely due to a combination of lower gas surface densities and higher gas velocity dispersions at lower stellar mass, which act to reduce gas fragmentation and lead to lower values of stellar clumpiness.   

    \item The decrease in concentration at lower stellar masses leads to a loss in the leverage that this parameter provides in separating early type and late type galaxies. As a result it becomes more difficult to separate these classes using morphological parameters as stellar mass decreases. 

    \item In the dwarf regime, galaxies with higher concentration, lower asymmetry and lower clumpiness have a higher probability of being red (i.e. quenched of star formation). These trends can be explained by the fact that higher concentrations tend to align with more pronounced bulges, which can induce stronger AGN feedback and morphological quenching. In a similar vein, a lower clumpiness is likely to be correlated with a lower gas surface density and higher velocity dispersion in gas disks, both of which would act to aid quenching. {\color{black}The morphological trends} are similar in massive galaxies, although the anti-correlation between clumpiness and redness appears to break down in this regime. 

    \item The incidence of bars declines with decreasing stellar mass, until it becomes consistent with zero around $M_{\star}$ $\sim$ 10$^{8}$ M$_{\odot}$, suggesting a lower limit to the stellar mass that is needed to induce bar formation. {\color{black}This decrease may be driven by the disks of lower stellar mass galaxies being more fragile and hot which, in turn, inhibits bar formation.}  
    
\end{itemize}

While our results are novel in their own right, our study demonstrates the science that is possible in the dwarf regime when a combination of deep photometric data and high resolution imaging is available. As such, our work offers a preview of what can be achieved by combining next-generation surveys like LSST and \textit{Euclid} in the near future, which are likely to result in significant advances in our understanding of galaxy evolution. 


\section*{Acknowledgements}

We are grateful to the anonymous referee for manu constructive comments which enables us to improve the quality of the original manuscript. SK, IL and AEW acknowledge support from the STFC (grant numbers ST/Y001257/1 and ST/X001318/1). SK also acknowledges a Senior Research Fellowship from Worcester College Oxford. We acknowledge support from the ERC Advanced Investigator Grant EPOCHS (788113) and support from the University of Manchester in the form of a studentship to LW.

{\color{black}This study has used the following online tool to measure the sizes of structures in galaxies: https://imagemeasurementonline.com/.} 

The Hyper Suprime-Cam (HSC) collaboration includes the astronomical communities of Japan and Taiwan, and Princeton University. The HSC instrumentation and software were developed by the National Astronomical Observatory of Japan (NAOJ), the Kavli Institute for the Physics and Mathematics of the Universe (Kavli IPMU), the University of Tokyo, the High Energy Accelerator Research Organization (KEK), the Academia Sinica Institute for Astronomy and Astrophysics in Taiwan (ASIAA), and Princeton University. Funding was contributed by the FIRST program from the Japanese Cabinet Office, the Ministry of Education, Culture, Sports, Science and Technology (MEXT), the Japan Society for the Promotion of Science (JSPS), Japan Science and Technology Agency (JST), the Toray Science Foundation, NAOJ, Kavli IPMU, KEK, ASIAA, and Princeton University. This paper makes use of software developed for Vera C. Rubin Observatory. We thank the Rubin Observatory for making their code available as free software at http://pipelines.lsst.io/.

This paper is based on data collected at the Subaru Telescope and retrieved from the HSC data archive system, which is operated by the Subaru Telescope and Astronomy Data Center (ADC) at NAOJ. Data analysis was in part carried out with the cooperation of Center for Computational Astrophysics (CfCA), NAOJ. We are honored and grateful for the opportunity of observing the Universe from Maunakea, which has the cultural, historical and natural significance in Hawaii. This paper used data that is based on observations collected at the European Southern Observatory under
ESO programme ID 179.A-2005 and on data products produced by CALET and
the Cambridge Astronomy Survey Unit on behalf of the UltraVISTA consortium.


\section*{Data Availability}

The observational data used in this study are taken from \citet{Weaver2022}. The JWST data used in this work are available in the COSMOS Web survey dataset, through the Mikulski Archive for Space Telescopes (https://mast.stsci.edu/). Additional data products will be shared on reasonable request to the first author. 


\bibliographystyle{mnras}
\bibliography{references} 


\appendix

\section{Morphological trends with stellar mass when only the bluest dwarfs are considered}

\label{app:bluest_dwarfs}

{\color{black}As noted in the introduction, shallow surveys like the SDSS only contain the bluest dwarf galaxies.} Here, we present a version of the analysis in Section \ref{sec:morphology_mass} where we only consider the percentiles in rest-frame colour, in the dwarf regime, that are likely to be detectable in the SDSS. 

To do this we consider the dwarf galaxy population in the NASA Sloan Atlas (NSA), which offers a catalogue of the properties of nearby galaxies in the SDSS at $z < 0.055$ \citep{Blanton2011}. A key feature of the NSA is its reprocessing of SDSS images using improved background subtraction that makes it more likely to accurately capture the properties of faint structures like dwarf galaxies and the outer regions of massive galaxies. 

To estimate the percentiles in colour that are likely to be detectable in the SDSS in the dwarf regime, we first calculate the number of dwarf galaxies in COSMOS2020 at 10$^{7}$ M$_{\odot}$ < $M_{\star}$ < 10$^{9.5}$ M$_{\odot}$ at $z<0.055$. We then multiply this value by the ratio of the sky areas of the NSA and COSMOS2020. This provides an estimate of the number of dwarf galaxies that would have appeared in the SDSS if it were as deep as COSMOS2020. Finally, we calculate the ratio between the actual number of dwarf galaxies that are visible in the SDSS to the estimate derived above. This ratio, around 3 per cent, represents an estimate of the subset of the true dwarf population that is visible in SDSS. In other words, the SDSS is likely to reveal only around the bluest 3 per cent of dwarfs in the nearby Universe.

\begin{figure}
\includegraphics[width=0.95\columnwidth]{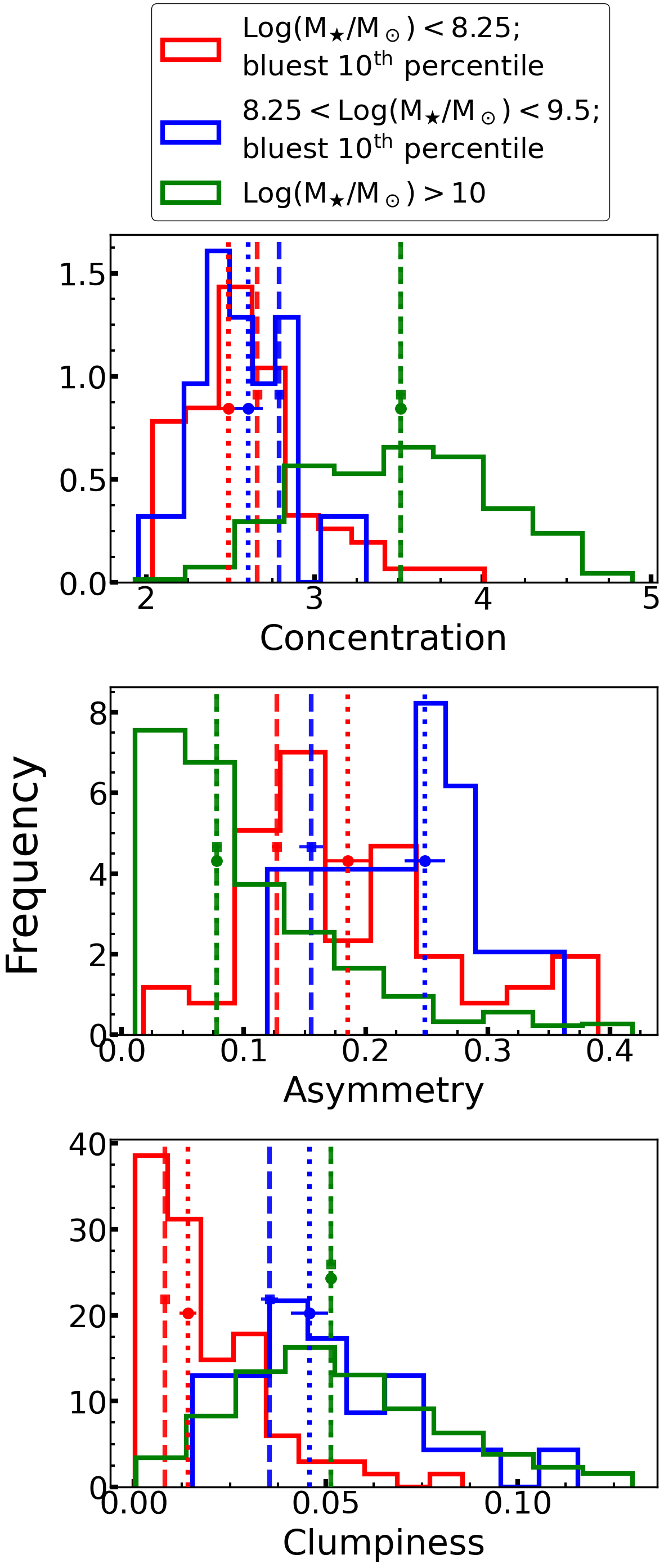}
\caption{The concentration (top), asymmetry (middle) and clumpiness (bottom) distributions as a function of stellar mass where for each dwarf stellar mass bin we use the galaxies in the bluest 10 per cent of the dwarf galaxy population. The red and blue histograms show the lower (10$^7$ M$_{\odot}$ < M$_{\star}$ < 10$^{8.25}$ M$_{\odot}$) and upper (10$^{8.25}$ M$_{\odot}$ < M$_{\star}$ < 10$^{9.5}$ M$_{\odot}$) halves of the stellar mass range of our dwarf population respectively, while the green histograms represent massive galaxies (M$_{\star}$ < 10$^{10}$ M$_{\odot}$). For each distribution we show the median value as a vertical dotted line with an associated error calculated via standard bootstrapping. The dashed vertical lines represent the medians of the full distributions from Figure \ref{fig:CAS_mass}.}
\label{fig:CAS_mass_bluest_dwarfs}
\end{figure}

Ideally we would like to explore how the analysis in Section \ref{sec:morphology_mass} changes if only the bluest 3 per cent of the dwarf population were considered. However, restricting our analysis to the bluest 3 per cent is not possible because the sample size becomes too small. We therefore consider how the findings of Section \ref{sec:morphology_mass} are altered if we consider only the bluest 10 per cent of dwarf population instead, noting that any differences are likely to be accentuated if we had used the bluest 3 per cent of dwarf galaxies. Table \ref{tab:blue_spearman} and Figure \ref{fig:CAS_mass_bluest_dwarfs} describe how Table \ref{tab:spearman} and Figure \ref{fig:CAS_mass} are modified if this subset of dwarfs is used. It is apparent that using a blue subset of dwarfs dilutes the trends between morphology and stellar mass, primarily because dwarfs in this subset have larger values of both asymmetries and clumpiness. 

\begin{table}
\begin{center}
\begin{tabular}{ c | c }
\toprule
& Rank correlation coefficient\\
\midrule
C vs Log M$_{\star}$ & +0.57\\
A vs Log M$_{\star}$ & -0.29\\
S vs Log M$_{\star}$ & +0.48
\end{tabular}
\caption{Results of Spearman's rank correlation tests between the CAS parameters and stellar mass when only the bluest 10 per cent of dwarfs are considered.} 
\label{tab:blue_spearman}
\end{center}
\end{table}

\section{The impact of uncertainties on the CAS trends in the effective surface brightness vs stellar mass plane}

\label{app:CAS_errors}

\begin{figure}
\includegraphics[width=0.95\columnwidth]{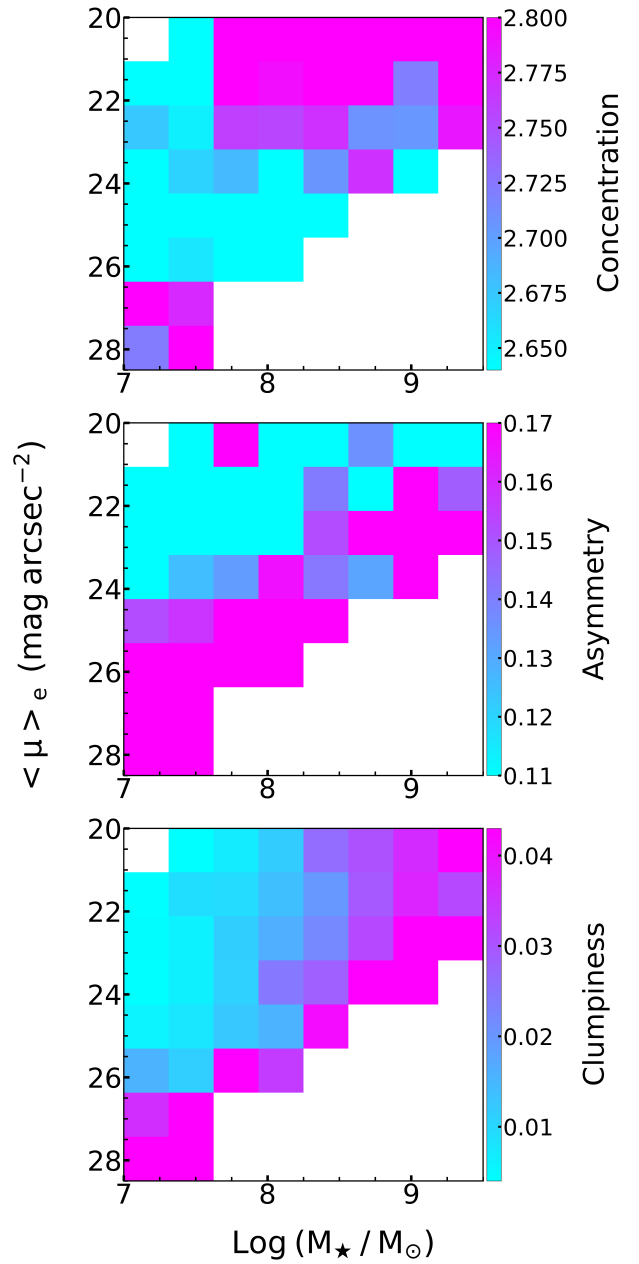}
\caption{The concentration (top), asymmetry (middle) and clumpiness (bottom) distributions in the effective surface brightness -- stellar mass plane, when the CAS values are resampled using their uncertainties.} 
\label{fig:CAS_effSB_mass_errors}
\end{figure}

In this section, we show that the CAS trends seen in the $\langle$$\mu_{\rm e}$$\rangle$ -- M$_\star$ plane in Section \ref{sec:morphology_mass} do not change when we resample the CAS values using their errors. {\color{black}Figure \ref{fig:CAS_effSB_mass_errors} shows an example of a scenario where the CAS values are shifted randomly to either their +1$\sigma$ or -1$\sigma$ positions and the heatmap is regenerated. All such scenarios produce the same result i.e. that the trends seen in Figure \ref{fig:CAS_effSB_mass} remain unaltered. These trends are therefore real and do not get washed out when the uncertainties on the parameters are considered.} 

\section{The effective surface brightness vs stellar mass plane extended to the massive galaxy regime}

\label{app:CAS_eff_massive}

\begin{figure*}
\includegraphics[width=0.95\textwidth]{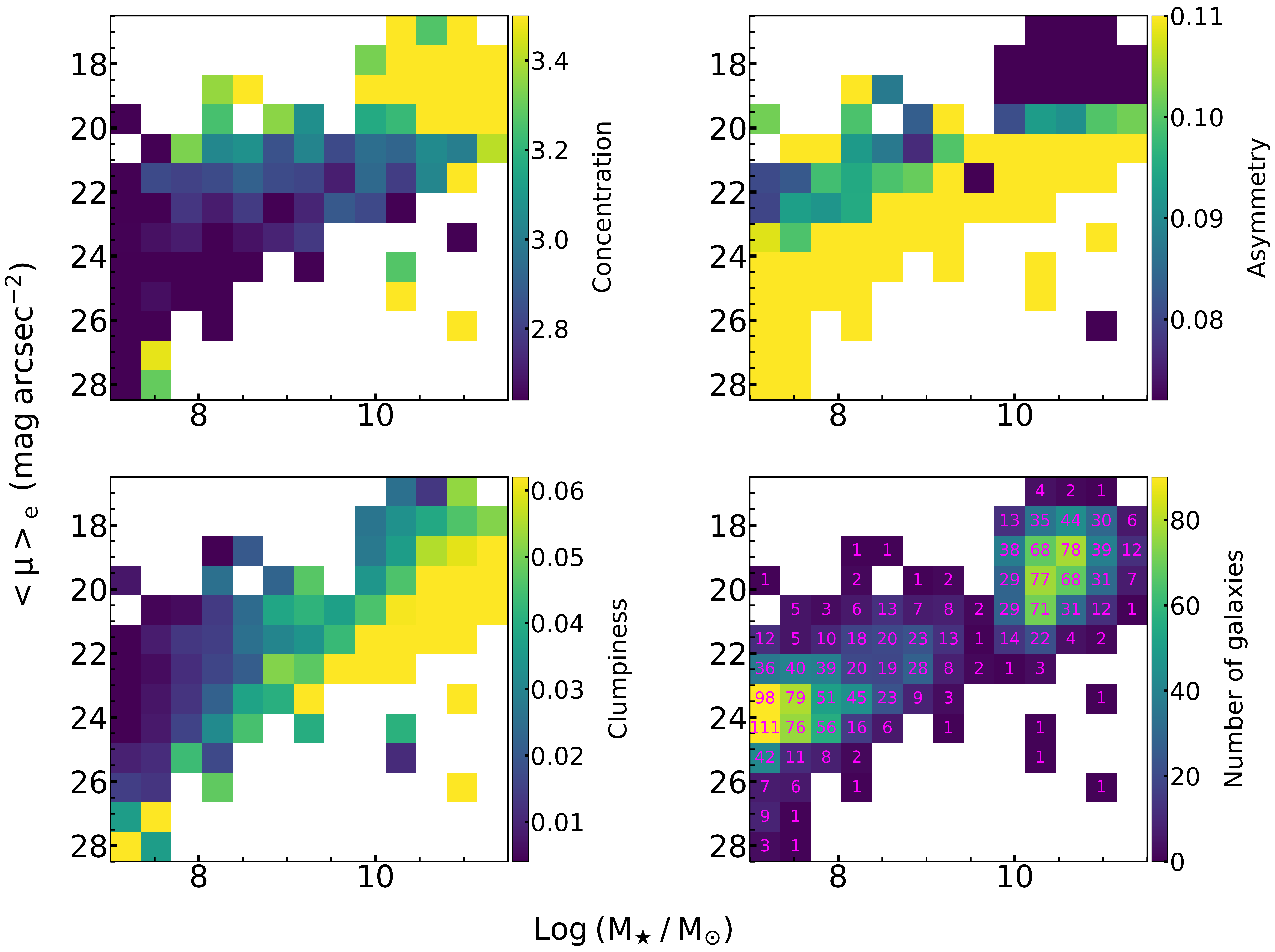}
\caption{The concentration (top-left), asymmetry (top-right) and clumpiness (bottom-left) distributions in the effective surface brightness -- stellar mass plane extended to the massive galaxy regime. The bottom-right panel shows the number of galaxies in different regions of this plane.}
\label{fig:CAS_effSB_mass_massive}
\end{figure*}

In this section, we extend the analysis of the $\langle$$\mu_{\rm e}$$\rangle$ -- M$_\star$ plane, colour coded by the CAS parameters, to the massive galaxy regime. Figure \ref{fig:CAS_effSB_mass_massive} indicates that the trends seen in the dwarf regime do not change when we extend our analysis to the massive regime. Since the range of the CAS parameters is greater when the massive and dwarf regimes are considered at the same time, the trends that are clearly seen in the dwarf population in Figure \ref{fig:CAS_effSB_mass} become less visible.  

\section{Examples of edge-on galaxies}

\label{app:edge_on_galaxies}

\begin{figure*}
\includegraphics[width=0.95\textwidth]{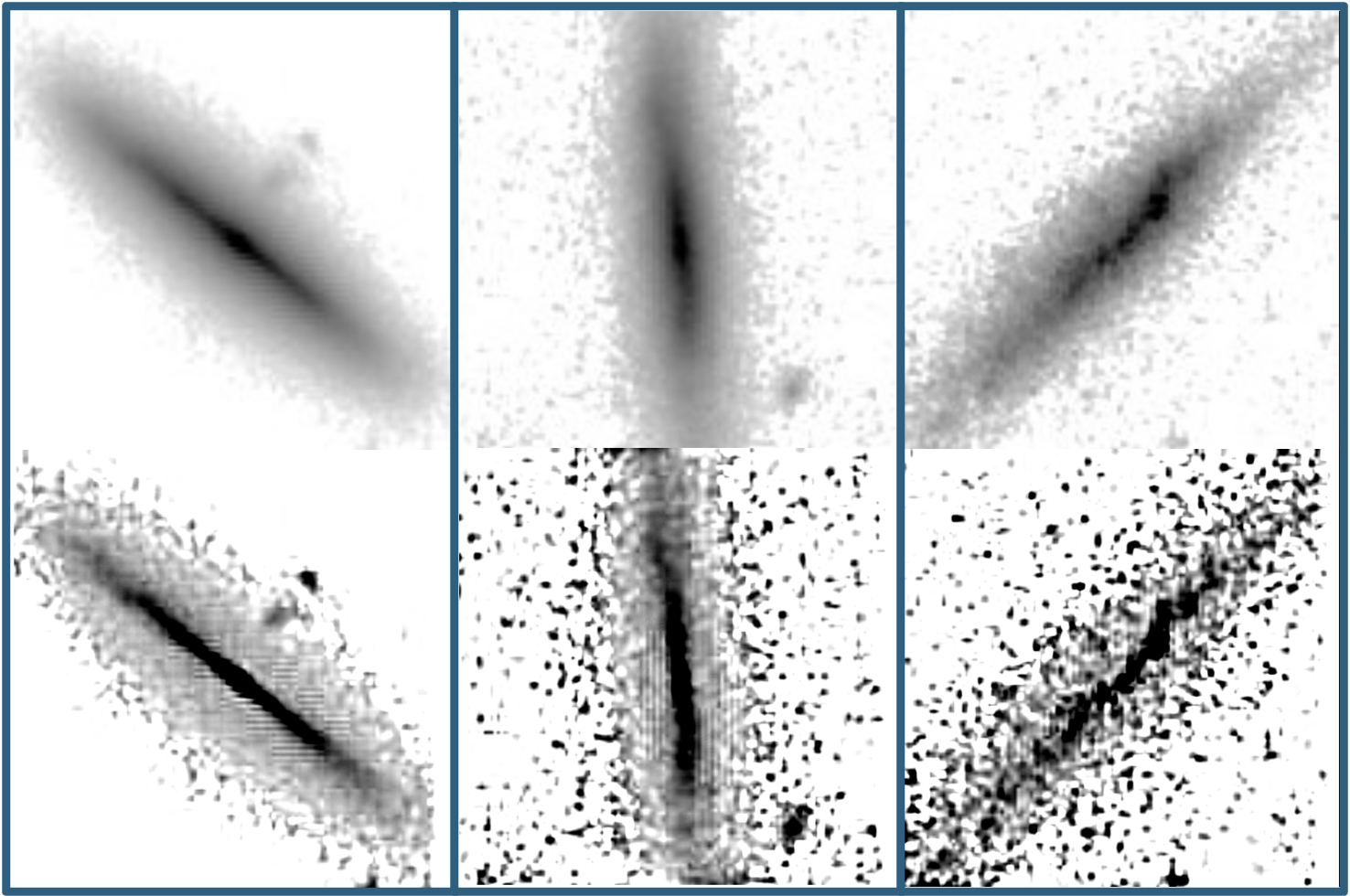}
\caption{Examples of edge-on systems in our galaxy sample. Each column shows an individual galaxy. The top panel presents the JWST F277W image, while the bottom panel shows its unsharp-masked counterpart. The size of each image is 5 arcseconds on a side.}
\label{fig:edge_on_examples}
\end{figure*}

{\color{black}Our estimate of the bar fraction excludes edge-on galaxies because bars cannot be identified in such systems. Figure \ref{fig:edge_on_examples} shows examples of such edge-on galaxies in our sample.}

\bsp
\label{lastpage}
\end{document}